\documentclass[sigconf]{acmart}
 \pdfoutput=1 
\usepackage{booktabs} 

\usepackage{algorithm,algpseudocode}
\usepackage[flushleft]{threeparttable}

\usepackage{textcomp}
\usepackage{xcolor,colortbl}

\usepackage{pgfplots}
\usepackage{tikz}
\usepackage{subcaption}
\usetikzlibrary{patterns}
\usepgfplotslibrary{groupplots}
\usepackage{caption} 
\usepackage{multirow}
\usepackage{graphicx}

\copyrightyear{2018} 
\acmYear{2018} 
\setcopyright{acmcopyright}
\acmConference[HT '18]{29th ACM Conference on Hypertext and Social Media}{July 9--12, 2018}{Baltimore, MD, USA}
\acmBooktitle{HT '18: 29th ACM Conference on Hypertext and Social Media, July 9--12, 2018, Baltimore, MD, USA}
\acmPrice{15.00}
\acmDOI{10.1145/3209542.3209565}
\acmISBN{978-1-4503-5427-1/18/07}

\begin{document}
\title{Sentiment-driven Community Profiling and Detection\\ on Social Media}

\author{Amin Salehi, Mert Ozer, Hasan Davulcu}
\affiliation{%
	\institution{ Computer Science and Engineering \\
		Arizona State University, Tempe, AZ, USA}
}
\email{{asalehi1, mozer, hdavulcu}@asu.edu}

\begin{abstract}
Web 2.0 helps to expand the range and depth of conversation on many issues and facilitates the formation of online communities. Online communities draw various individuals together based on their common opinions on a core set of issues. Most existing community detection methods merely focus on discovering communities without providing any insight regarding the collective opinions of community members and the motives behind the formation of communities. Several efforts have been made to tackle this problem by presenting a set of keywords as a community profile. However, they neglect the positions of community members towards keywords, which play an important role for understanding communities in the highly polarized atmosphere of social media. To this end, we present a sentiment-driven community profiling and detection framework which aims to provide community profiles presenting positive and negative collective opinions of community members separately. With this regard, our framework initially extracts key expressions in users' messages as representative of issues and then identifies users' positive/negative attitudes towards these key expressions. Next, it uncovers a low-dimensional latent space in order to cluster users according to their opinions and social interactions (i.e., retweets). We demonstrate the effectiveness of our framework through quantitative and qualitative evaluations.
\end{abstract}

%
%
%


\maketitle

\section{Introduction}

With the advent of social media platforms, individuals are able to express their opinions on a variety of issues online. Like-minded users forge online communities by interacting with each other and expressing similar attitudes towards a set of issues. While many methods \cite{papadopoulos2012community} have been proposed to detect online communities, most of them do not provide insights into the collective opinions of community members. To shed light on such opinions, few efforts have focused on profiling communities, but a large body of work has been devoted to user profiling \cite{mislove2010you,harvey2013building,ikeda2013twitter}. Indeed, ``the founders of sociology claimed that the causes of social phenomena were to be found by studying groups rather than individuals'' \cite{hechter1988principles}.

Turner et al. \cite{turner1987rediscovering} suggest that individuals come together and form communities by developing shared social categorization of themselves in contrast to others . Therefore, to profile a community, we need to uncover the collective opinions of its members which makes them distinguishable from the members of other communities. Tajfel \cite{tajfel2010social} suggests focusing on unit-forming factors (e.g., similarities, shared threats, or common fate) which function as cognitive criteria for segmentation of the social world into discrete categories. Accordingly, the controversial issues on which users have different opinions can be taken into account in order to discover the motives driving the segmentation of social media and the formation of communities. As a result, the profile of a community should present its important issues on which its members generally have the same position. Such community profiles can be found useful in a broad range of applications such as recommender systems \cite{sahebi2011community}, community ranking \cite{chen2008combinational, han2016csd}, online marketing \cite{kozinets2002field}, interest shift tracking of communities \cite{zhou2012community}, and community visualization \cite{cruz2013community}. For example, a group recommender system \cite{boratto2016group} can suggest more relevant items to communities by knowing the collective opinions of their members.

Many community detection methods \cite{cai2017community,zhou2012community,akbari2017leveraging,natarajan2013community,ozer2016community,pathak2008social,pei2015nonnegative,sachan2012using,zhou2006probabilistic} which are capable of community profiling have been proposed. However, these methods usually present a set of frequent keywords used by the members of a community as the community profile. However, it is common in social media that the members of different communities use the same keywords in their messages. Therefore, keywords alone might not be enough to differentiate communities in which their members have similar word usage. For instance, in the course of the US presidential election of 2016, Republicans and Democrats have used many common keywords such as Trump, Clinton, and Obamacare but with different sentiments. To differentiate and understand these two parties, not only keywords but also the collective attitude of community members towards these keywords should be taken into account.



In this paper, we tackle the aforementioned problem by proposing a sentiment-driven community profiling and detection framework which utilizes user-generated content and social interactions. Our framework first captures key expressions in users' messages as representative of issues by utilizing a POS-tagger and built-in features of social media platforms (i.e. hashtags and user accounts). Next, it identifies users' attitudes towards the extracted key expressions. Finally, we employ a novel graph regularized semi-nonnegative matrix factorization (GSNMF) technique to cluster users according to both their opinions and social interactions. GSNMF uncovers not only communities but also their sentiment-driven profiles. The main contributions of the paper are as follows:
\begin{itemize}
\item Providing sentiment-driven community profiles which separately present the positive and negative collective attitudes of the members of each community towards their important key expressions;
\item Achieving higher performance in detecting communities compared to several existing state-of-the-art community detection methods.
\end{itemize}

The rest of the paper is organized as follows. We review related work in Section 2. In Section 3, we propose our sentiment-driven community profiling and detection framework. To demonstrate the efficacy of our framework, we conduct quantitative and qualitative experiments by using real-world social media datasets in Section 4. Section 5 concludes the paper and discusses future work.

\section{Related Work}
Community detection methods can fall into three broad categories: link-based, content-based and hybrid methods. Most of the existing works belong to the first category and utilize only social interactions \cite{clauset2004finding,blondel2008fast}. However, they neglect to utilize valuable user-generated content in which users express their opinions. On the other hand, content-based methods only utilize user-generated content \cite{lee2013campaign}. Nevertheless, the content on social media is extremely noisy, resulting in the failure in detecting communities effectively. To alleviate these challenges, hybrid community detection methods are proposed. These methods are the most related work to our study since they not only exploit both user-generated content and social interactions but are also capable of profiling communities. These methods roughly fall into two categories: probabilistic graphical models and non-negative matrix factorization (NMF) based methods.


\subsection{Probabilistic Graphical Models}
Community User Topic (CUT) models \cite{zhou2006probabilistic} are
one of the earliest works for detecting communities using probabilistic graphical models. The first proposed model (CUT\textsubscript{1}) assumes that a community is a distribution over users, while the second one (CUT\textsubscript{2}) considers a community as a distribution over topics. To discover communities, CUT\textsubscript{1} and CUT\textsubscript{2} are biased towards social interactions and user-generated content, respectively. Community Author Recipient Topic (CART) \cite{pathak2008social} is an unbiased model which assumes the members of a community discuss topics of mutual interests and interact with one other based on these topics. CART considers users as both authors and recipients of a message. However, in well-known social networks such as Twitter and Facebook, the number of recipients for a message can be very large. To make community detection scalable, Topic User Community Model (TUCM) \cite{sachan2012using}, considering users as authors not recipients, is proposed. Since CART and TUCM consider users as authors, recipients, or both, they are limited to certain types of social interactions (e.g., retweet and reply-to in Twitter). The link-content model \cite{natarajan2013community} solves this problem by ignoring the assumption that messages can be related to each other using social interactions. It is also capable of using different types of social interactions (e.g., friendship in Facebook and followership in Twitter). Furthermore, COCOMP \cite{zhou2012community} is proposed to model each community as a mixture of topics about which a corresponding group of users communicate. \cite{cai2017community} is another model which detects and profiles communities in the domains having user-user, user-document, and document-document links.

\begin{table}
	\centering
	\caption{Notations used in the paper}
	\label{tab:notation}
	\small
	\begin{tabular}{|c|l|} \hline
		Notation & Explanation\\ \hline
		$\mathcal{U}$ & The set of users\\ \hline
		$\mathcal{C}$ & The set of communities\\ \hline
		$\mathcal{S}$ & The set of key expressions\\ \hline
		$n$ & The number of users\\ \hline
		$m$ & The number of key expressions\\ \hline
		$k$ & The number of communities\\ \hline
		$\mathbf{X}$ & User opinion matrix\\ \hline
		$\mathbf{U}$ & Community membership matrix\\ \hline
		$\mathbf{V}$ & Community profile matrix\\ \hline
		$\mathbf{W}$ & Social Interaction matrix\\ \hline
		$\overset{\sim}{\mathbf{W}}$ & Symmetrically normalized matrix $\mathbf{W}$\\ \hline
		$\mathbf{D}$ & Degree matrix of $\mathbf{W}$\\ \hline
	\end{tabular}
\end{table}

\subsection{NMF-based Methods}
In order to encode graphs as local geometric structures, many methods extending standard NMF are proposed. LLNMF \cite{gu2009local} introduces a regularizer, imposing the constraint that each data point should be clustered based on the labels of the data points in its neighborhood. GNMF \cite{cai2011graph} further incorporates a graph regularizer to encode the manifold structure. Moreover, DNMF \cite{shang2012graph} is proposed based on the the idea that not only the data, but also the features lie on a manifold.
The graph regularizers proposed by the above methods have been utilized by several other works \cite{pei2015nonnegative,ozer2016community} to detect communities on social media. Moreover, another work \cite{akbari2017leveraging} proposes a NMF-based approach utilizing a graph regularizer to exploit different social views (i.e., different social interactions and user-generated content) as well as prior knowledge in order to detect and profile communities.

\section{The Proposed Framework}
\subsection{Problem Statement}
We first begin with the introduction of the notations used in the paper as summarized in Table \ref{tab:notation}. 
Let $\mathcal{U}=\{u_1,u_2,...,u_n\}$ be the set of $n$ users, $\mathcal{C}=\{c_1,c_2,...,c_k\}$ indicate the set of $k$ communities, and $\mathcal{S}=\{s_1,s_2,...,s_k\}$ denote the set of $m$ key expressions. $\mathbf{X} \in \mathbb{R}^{m \times n} $ indicates the matrix of users' attitudes towards key expressions, where $\mathbf{X}_{li}$ corresponds to the attitude of user $u_i$ towards key expression $s_l$. Furthermore, $\mathbf{U} \in \mathbb{R}_+^{n \times k} $ indicates the community membership matrix,  in which $\mathbf{U}_{ik}$ corresponds to the membership strength of user $\mathbf{U}_i$ in community $c_k$. $\mathbf{V} \in \mathbb{R}^{m \times k}$ further denotes the community profile matrix, where  $\mathbf{V}_{lk}$ corresponds to the contribution strength of key expression $s_l$ in the profile of community $c_k$. Moreover, $\mathbf{W} \in \mathbb{R}_+^{n \times n}$ indicates the social interaction matrix, in which $\mathbf{W}_{ij}$ represents the number of social interactions between user $u_i$ and user $u_j$.  We use $\overset{\sim}{\mathbf{W}}$ to denote the symmetric normalization of $\mathbf{W}$ (i.e., $\overset{\sim}{\mathbf{W}}=\mathbf{D}^{-1/2} \mathbf{W} \mathbf{D}^{-1/2}$, where $\mathbf{D}$ is the degree matrix of $\mathbf{W}$). 

By using the above notations, the problem of detecting and profiling communities can be defined as: \textit{Given user opinion matrix $\mathbf{X}$ and social interaction matrix $\mathbf{W}$ , we aim to obtain community membership matrix $\mathbf{U}$ and community profile matrix $\mathbf{V}$.}

\subsection{Extracting Key Expressions as Issues}
Social media presents an opportunity to utilize user-generated content in which individuals express their opinions on various issues. The first step towards understanding users' opinions is the extraction of the issues they discuss. To this end, many efforts \cite{qiu2011opinion, mukherjee2012aspect, zhang2014aspect} have been made to extract issues or related aspects. However, these methods require enough training samples for a specific domain to work accurately. Due to the lack of such a dataset for our required experiments, we follow a simple approach to extract key expressions. We utilize the  built-in features common among well-known social media platforms. In such social networks, hashtags and user account mentions, which usually indicate issues, are perpended by '\#' and '@', respectively. However, the built-in features are not enough to detect all issues. To tackle this problem, we employ a part-of-speech (POS) tagger to extract proper nouns and noun phrases (two or more nouns in a row) as representative of issues. If some proper nouns  are in a row, they are considered as a single key expression. We utilize the POS tagger proposed in \cite{gimpel2011part} proven to perform well for the content on social media.

\subsection{Capturing Users' Opinions}
The position individuals take towards issues reflects their opinions\footnote{An opinion is defined as an attitude towards an issue \cite{fishbein1977belief}.}. Many efforts \cite{pontiki2016semeval, tang2016aspect} have been made to detect users' sentiments towards issues. However, these methods work effectively when enough training samples for a specific domain are given. However, there is no such a dataset for our required experiments so we apply a simple approach although a sophisticated approach can improve the result of our framework. First, a window with a certain size centered at each positive/negative sentiment word is created. Next, the nearest key expression to the sentiment word is selected, and the positivity/negativity of the sentiment word determines user's positive/negative attitude towards that key expression. For instance, in the message "Conservatives seem angry every time economy adds jobs", we assume the author has a negative sentiment towards key expression "conservatives" because it is the closest key expression to the negative sentiment word "angry" if we consider the window size to be at least two. To generate matrix $\mathbf{X}$, we need to apply the above procedure for all messages. Therefore, for each message if author $u_i$ takes a positive/negative attitude towards key expression $s_l$, we add the sentiment strength of the corresponding sentiment word to $\mathbf{X}_{li}$, respectively. We utilize SentiStrength \cite{thelwall2010sentiment} to discover positive and negative words as well as their sentiment strength. 




\subsection{Exploiting Users' Opinions}
After extracting users' attitudes towards key expressions, the next major objective is sentiment-driven community profiling and detection of like-minded users. To accomplish this, we exploit semi-nonnegative matrix factorization \cite{ding2010convex} as follows:
\begin{equation}
\label{proposed_text}
\begin{aligned}
\underset{U,V}{\text{min}}\quad & ||\mathbf{X} - \mathbf{V} \mathbf{U}^T ||_F^2 \\
\text{s.t.}\quad & \mathbf{U} \geq 0.
\end{aligned}
\end{equation}

Since the non-negativity constraint in Eq. \eqref{proposed_text} only holds on matrix $\mathbf{U}$, matrix $\mathbf{V}$ can contain both positive and negative values. A positive/negative value of $\mathbf{V}_{lk}$ denotes that the members of community $c_k$ have a collective positive/negative attitude towards key expression $s_l$. The larger the positive value of $\mathbf{V}_{lk}$ is, the more the members of community $c_k$ have a collective positive attitude towards key expression $s_l$. The lower the negative value of $\mathbf{V}_{lk}$ is, the more the members of community $c_k$ have a collective negative attitude towards key expression $s_l$. This property of matrix  $\mathbf{V}$ also results in the categorization of key expressions into positive and negative categories according to the sign of the corresponding elements of key expressions in matrix $\mathbf{V}$. Therefore,
key expressions in a community profile are divided into two positive and negative categories. Moreover, the key expressions in each category can also be ranked by their values in matrix $\mathbf{V}$ in order to show how important they are to the members of the corresponding community.

\subsection{Exploiting Social Interactions}
Social interactions (e.g., retweets in Twitter and friendship in Facebook) are one of the most effective sources of information to detect communities \cite{papadopoulos2012community}. To utilize social interactions, NMF-based methods exploit graph regularizers. Gu \textit{et al.} \cite{gu2011trivial} suggest that graph regularizers used in GNMF \cite{cai2011graph} and DNMF \cite{shang2012graph} suffer from the trivial solution problem and the scale transfer problem. When the graph regularizer parameter is too large, the trivial solution problem occurs and results in similarity among the elements of each row of community membership matrix $\mathbf{U}$. The scale transfer problem, in which $\{ \mathbf{V}^*, \mathbf{U}^* \}$ stands as the optimal solution for Eq. \eqref{proposed_text}, results in a smaller objective value for the scaled transferred solution $(\frac{\mathbf{V}*}{\beta}, \beta \mathbf{U}')$ , for any real scalar $\beta>1$.

To avoid these problems, we propose using the following graph regularizer,
\begin{equation}
\label{proposed_graph-reg}
\begin{aligned}
\underset{U}{\text{max}}\quad & Tr(\mathbf{U}^T \overset{\sim}{\mathbf{W}} \mathbf{U}) \\
\text{s.t.}\quad & \mathbf{U} \geq 0, \mathbf{U}^T \mathbf{U} = \mathbf{I}.
\end{aligned}
\end{equation}

where $\mathbf{I}$ is the identity matrix with the proper size. Eq. \eqref{proposed_graph-reg} clusters users into $k$ communities, with the most interactions within each community and the fewest interactions between communities. In fact, Eq. \eqref{proposed_graph-reg} is equivalent to the nonnegative relaxed normalized cut as put forth in \cite{ding2005equivalence}.

\subsection{The Proposed Framework GSNMF}
In the previous sections, we introduced our solutions to exploit and social interactions and users' attitudes toward key expressions. Using these solutions, our proposed framework simultaneously utilizes users' opinions and social interactions to uncover communities and their profiles. The proposed framework requires solving the following optimization problem,

\begin{algorithm}[tbh]
	\caption{The Proposed Algorithm for GSNMF}\label{alg:1}
	\begin{algorithmic}[1]
		\Statex{\textbf{Input:} user opinion matrix $\mathbf{X}$ and social interaction matrix $\mathbf{W}$}
		\Statex{\textbf{output:} community membership matrix $\mathbf{U}$ and community profile matrix $\mathbf{V}$}
		\State {Initialize $\mathbf{U}$ and $\mathbf{V}$ randomly where $\mathbf{U}\geq 0$}
		\While{ not convergent}
		\State Update $\mathbf{U}$ according to Eq. \eqref{U_updating}
		\State Update $\mathbf{V}$ according to Eq. \eqref{V_updating}
		\EndWhile
	\end{algorithmic}
\end{algorithm}

\begin{equation}
\label{proposed_method}
\begin{aligned}
\underset{U,V}{\text{min}}\quad & \mathcal{F} = ||\mathbf{X} - \mathbf{V} \mathbf{U}^T ||_F^2 - \lambda Tr(\mathbf{U}^T \overset{\sim}{\mathbf{W}} \mathbf{U})\\
\text{s.t.}\quad & \mathbf{U} \geq 0, \mathbf{U}^T \mathbf{U} = \mathbf{I}.
\end{aligned}
\end{equation}

where $\lambda$ is a non-negative regularization parameter controlling the contribution of the graph regularizer in the final solution. Since the optimization problem in Eq. \eqref{proposed_method} is not convex with respect to variables $\mathbf{U}$ and $\mathbf{V}$ together, there is no guarantee to find the global optimal solution. As suggested by \cite{lee2001algorithms}, we introduce an alternative scheme to find a local optimal solution to the optimization problem. The key idea is optimizing the objective function with respect to one of the variables $\mathbf{U}$ or $\mathbf{V}$, while fixing the other one. The algorithm keeps updating the variables until convergence.\\

Optimizing the objective function $\mathcal{F}$  with respect to $\mathbf{U}$ leads to the following update rule,

\begin{equation}
\label{U_updating}
\begin{aligned}
\mathbf{U} = \mathbf{U} \odot \sqrt{\frac{(\mathbf{X}^T \mathbf{V})^+ + [\mathbf{U}(\mathbf{V}^T \mathbf{V})^-] + \lambda \overset{\sim}{\mathbf{W}} \mathbf{U} + \mathbf{U} \mathbf{\Gamma}^-} {(\mathbf{X}^T \mathbf{V})^- + [\mathbf{U}(\mathbf{V}^T \mathbf{V})^+] + \mathbf{U} \mathbf{\Gamma}^+}}
\end{aligned}
\end{equation}

where $\odot$ denotes the Hadamard product, and
\begin{equation}
\begin{aligned}
\mathbf{\Gamma} = \mathbf{U}^T \mathbf{X}^T \mathbf{V} - \mathbf{V}^T \mathbf{V} + \lambda \mathbf{U}^T \overset{\sim}{\mathbf{W}} U
\end{aligned}
\end{equation}

We separate the negative and positive parts of a matrix $\mathbf{A}$ as $\mathbf{A}^-=(|\mathbf{A}|-\mathbf{A})/2$ and $\mathbf{A}^+ = (|\mathbf{A}|+\mathbf{A})/2$, respectively. The details regarding the computation of Eq \eqref{U_updating} are given in Appendix A.

Moreover, optimizing the objective function $\mathcal{F}$  with respect to $\mathbf{V}$ leads to the following updating rule,

\begin{equation}
\label{V_updating}
\begin{aligned}
\mathbf{V} = \mathbf{X} \mathbf{U} (\mathbf{U}^T\mathbf{U})^{-1}
\end{aligned}
\end{equation}

\noindent The details are given in Appendix B. 

The algorithm for GSNMF is shown in Algorithm \ref{alg:1}. In line 1, it randomly initializes $\mathbf{U}$ and $\mathbf{V}$. From lines 2 to 5, it updates $\mathbf{U}$ and $\mathbf{V}$ until convergence is achieved.

\subsection{Algorithm Complexity}
In Algorithm \ref{alg:1}, the most costly operations are the matrix multiplications in update rules Eq. \eqref{U_updating} and Eq. \eqref{V_updating}. Therefore, we provide the time complexity of these two updating rules as follows:

\begin{itemize}
	\item The time complexity of Eq. \eqref{U_updating} is $O(nmk+mk^2+n^2k+nk^2)$.
	\item Since the inversion of small matrix $\mathbf{U}^T\mathbf{U}$ is trivial, the time complexity of Eq. \eqref{V_updating} is $O(mnk+nk^2)$.
\end{itemize}

Accordingly, the time complexity of Algorithm \ref{alg:1} is $O(ik(nm+mk+n^2+nk))$ where $i$ is the number of iterations. Our framework can be applied to large scale social network platforms by exploiting the distributed approaches outlined in \cite{liu2010distributed,gemulla2011large,li2014fast}.

\section{Experiments}
To evaluate the efficacy of our framework, we need to answer the following two questions:
\begin{enumerate}
	\item How effective is the proposed framework in detecting communities compared to the the-state-of-the-art community detection methods?
	\item How effective is our framework in profiling communities?
\end{enumerate}

In the next sections, we first describe the datasets used in this study. Next, the performance of GSNMF is compared with several state-of-the-art community detection methods. Then, we qualitatively evaluate the community profiles uncovered by our framework.

\begin{table}
	\centering
	\caption{The statistics of the datasets.}
	\label{tab:statistics}
	
	\resizebox{\columnwidth}{!}{%
		\begin{tabular}{l c c c}
			\hline
			& US & UK & Canada\\ \hline \hline
			\# of tweets & 113,818 & 236,008 & 98,899 \\
			\# of retweets & 18,891 & 6,863 & 3,104 \\
			\# of distinct words & 5,773 & 7,653 & 3,738 \\
			\# of distinct key expressions & 165 & 349 & 69 \\
			\# of users & 404 & 317 & 102 \\
			\# of baseline communities & 2 & 5 & 3 \\
			\hline
		\end{tabular}
	}
\end{table}

\subsection{Data Description}
We take politics as an example to evaluate our framework. In this regard, we used Twitter search API to crawl politicians' tweets from three different countries, namely United States, United Kingdom, and Canada. However, Twitter API imposes the limitation of retrieving only the latest 3200 tweets for each user. To overcome this limitation, we crawled politicians' user accounts several times during the time each dataset covers. The datasets are described as follows,
\begin{itemize}
	\item \textbf{US Dataset} consists of the tweets posted by 404 politicians from two major political parties (Republican party and Democratic party) in the United States from August 26 to November 29, 2016.
	\item \textbf{UK Dataset} consists of the tweets posted by 317 political figures from five major political parties (Conservative Party, Labour Party, Scottish National Party, Liberal Democratic Party, and UK Independence Party) in the United Kingdom from January 1 to September 30, 2015.
	\item \textbf{Canada Dataset} consists of the tweets posted by 102 politicians from three major political parties (Liberal Party, Conservative Party, and New Democratic Party) from January 1 to November 18, 2016.
\end{itemize}

All users in the datasets have discussed at least 15 key expressions. Moreover, the key expressions used by less than 15 users and stop words are eliminated. As a window size, we experimentally determine the threshold of 3 for the nearest keywords on both sides of each sentiment word. Furthermore, the party to which a user belong is labeled as ground truth. The statistics for the datasets are shown in Table
\ref{tab:statistics}. The GSNMF code and users' Twitter accounts as well as their ground truth labels used in this paper are available \footnote{\url{https://github.com/amin-salehi/GSNMF}}.

\subsection{Community Detection Evaluation}

\subsubsection{Baselines}
In order to demonstrate the effectiveness of our framework, we compare GSNMF with the following state-of-the-art community detection methods,
\begin{itemize}
	\item \textbf{GNMF} \cite{cai2011graph} is a hybrid method utilizing both user-generated and social interactions by incorporating a graph regularizer into standard NMF.
	\item \textbf{Louvain} \cite{blondel2008fast} is a link-based method optimizing modularity using a greedy approach.
	\item \textbf {Infomap} \cite{rosvall2008maps} is a link-based method built upon information theory to compress the description of random walks in order to find community structure.
	\item \textbf{DNMF} \cite{shang2012graph} is a hybrid method utilizing both user-generated content and social interactions by incorporating two regularizers (i.e, a graph regularizer and a word similarity regularizer) into standard NMF.
	\item \textbf{Soft Clustering} \cite{yu2005soft} is a link-based method that assigns users to communities in a probabilistic way.
	\item \textbf{CNM} \cite{clauset2004finding} is a link-based method based on modularity optimization.
\end{itemize}

\subsubsection{Evaluation Metrics}
To evaluate the performance of the methods, we utilize three metrics frequently used for community detection evaluation; namely, Normalized Mutual Information (NMI), Adjusted Rand Index (ARI), and purity.

\subsubsection{Experimental Results}
For this experiment, we use all three datasets. We also utilize the party membership of each politician as ground truth in our evaluation. For the methods providing soft community membership, like our framework, we select the community with the highest membership value for each user as the community to which she/he belongs.
Regularization parameters of NMF-based methods are set to be all powers of $10$ from $0$ to $9$ to find the best configuration for each of these methods. We run each method 10 times with its best configuration and then report the best result. According to the results shown in Table \ref{tab:election}, we can make the following observations,
\begin{itemize}
	\item Our proposed framework achieves the highest performance in terms of NMI and ARI for all three datasets. In terms of purity, it also achieves the best in the Canada and US datasets. In the UK dataset, Louvain, Infomap, and CNM obtain higher purity compared to our framework since they generate an artificially large number of communities for sparse graphs such as social media networks. For instance, Louvain detects 21 communities for UK dataset.
	\item Exploiting both user-generated content and social interactions does not necessarily result in achieving better performance compared to link-based methods. For example, the Soft Clustering method achieves better results compared to GNMF and DNMF in terms of all three used metrics. However, link-based methods do not uncover any community profile.
	\item All NMF-based methods achieve their highest performance with large values (i.e., from $10^6$ to $10^9$) for the graph regularizer parameter.
	
\end{itemize}

\subsection{Community Profiling Evaluation}
In this section, we evaluate the effectiveness of our proposed framework in profiling communities by using US and UK datasets. In this regard, we first label each community detected by our framework with the party to which the majority of community members belong. Next, we evaluate how effectively the profile of a community represents its corresponding ground truth party. To this end, two graduate students who have knowledge of US and UK politics are assigned to label the results of community profiling methods. It is asked that each key expression in a community profile to be assigned to one of the following categories:
\begin{itemize}
	\item Supported: A key expression is labeled as supported if the majority of community members have a positive attitude towards it or support it.
	\item Opposed: A key expression is labeled as opposed if the majority of community members have a negative attitude towards it or oppose it.
	\item Concerned: A key expression is labeled as concerned if the majority of community members are concerned about it.
	\item Unrelated: A key expression is labeled as unrelated if the annotators cannot find a strong relevance between the community (party) and the key expression.
\end{itemize}

In the tables representing community profiles, we color (and mark) supported, opposed, and concerned key expressions with green ($+$), red ($-$), and blue ($\pm$), respectively. We also leave unrelated key expressions uncolored (and unmarked).

In the following experiments, we expect our proposed framework to achieve three goals:
\begin{enumerate}
	\item Uncovering community profiles which represent the collective opinions of community members into two positive/negative categories;
	\item Assigning supported key expressions and opposed/concerned ones to positive/negative categories, respectively;
	\item Minimizing the number of unrelated key expressions in community profiles.
\end{enumerate}

In Sections \ref{US_Politics} and \ref{UK_Politics}, we evaluate the results of GSNMF according to the first and second goals by using US and UK datasets. To evaluate the performance of the third goal, Section \ref{Quantitative_results} compares GSNMF with the baselines with regard to their effectiveness in extracting relevant key expressions.

\begin{table}
	\centering
	\tiny
	\caption{Comparison of community detection methods.}
	\label{tab:election}
	\begin{tabular}{|l | c c c | c c c | c c c |}
		\hline
		\centering & \multicolumn{3}{|c|}{US} & \multicolumn{3}{|c|}{UK} & \multicolumn{3}{|c|}{Canada}  \\
		\hline
		Method & NMI & ARI & Purity & NMI & ARI & Purity & NMI & ARI & Purity \\ \hline \hline
		Louvain & 0.5083 & 0.3889 & 0.9752 & 0.7077 & 0.4352 & \textbf{0.9937} & 0.8602 & 0.8430 & 0.9902 \\
		Infomap & 0.5026 & 0.3755 & 0.9752 & 0.8871 & 0.8874 & 0.9936 & 0.8971 & 0.9299 & 0.9804 \\
		CNM & 0.5741 & 0.4664 & 0.9752 & 0.8830 & 0.8746 & 0.9905 & 0.9405 & 0.9643 & 0.9902 \\
		GNMF & 0.8564 & 0.9126 & 0.9777 & 0.8120 & 0.8291 & 0.9085  & 0.9597 & 0.9794 & 0.9902 \\
		DNMF & 0.8599 & 0.9222 & 0.9802 & 0.8308 & 0.8030 & 0.8896 & 0.9574 & 0.9716 & 0.9902 \\
		Soft Clustering & 0.8934 & 0.9413 & 0.9851 & 0.8481 & 0.8450 & 0.9495 & \textbf{1.0000} & \textbf{1.0000} & \textbf{1.0000}   \\
		GSNMF & \textbf{0.9069} & \textbf{0.9510} & \textbf{0.9876} & \textbf{0.9298} & \textbf{0.9612} & 0.9811  & \textbf{1.0000} & \textbf{1.0000} & \textbf{1.0000}\\
		\hline
	\end{tabular}
\end{table}

\subsubsection{US Politics}
\label{US_Politics}
The US dataset covers many events such as occurrences of gun violence, police brutality (e.g., the shooting of Terence Crutcher), the Flint water crisis, and the death of Fidel Castro; but the major event is the US presidential election of 2016. To give brief background knowledge, two major US parties during the election are described as follows \cite{lilleker2016us},
\begin{itemize}
	\item \textbf{Democratic Party:} A liberal party focusing on social justice issues. In 2016, Hillary Clinton was nominated as the presidential candidate of the party with Tim Kaine as her vice president. Moreover, Barrack Obama, the incumbent Democratic President, was a strong advocate for Hillary Clinton.
	\item \textbf{Republican Party:} A conservative party, known as the GOP, which had the majority of congressional seats in 2016 and embraces Judeo-Christian ethics. Moreover, Donald Trump was nominated as the party candidate for the presidency with Mike Pence as his vice president.
\end{itemize}

During the campaign, Republicans---especially Donald Trump---mainly criticized president Obama and his policies (e.g., Obamacare, tax plans, and Iran deal) in order to discredit Hillary Clinton, whom they claimed was going to continue the Obama legacy and uphold the status quo \cite{lilleker2016us}. On the other hand, Clinton's campaign brought the issue of gun violence into the contest, and also focused on human rights for groups such as women and LGBTQ \cite{lilleker2016us}.

\begin{table}
	\centering
	\tiny
	\caption{The profiles of two communities detected by our framework in the US dataset.}
	\label{tab:US-motives}
	\resizebox{\columnwidth}{!}{%
	\begin{threeparttable}
		\setlength\tabcolsep{1.5pt}
		\begin{tabular}{| c c | c c || c c | c c|}
			\hline
			\multicolumn{4}{|c||}{Democrats} & \multicolumn{4}{c|}{Republicans} \\ 
			\hline
			& Positive & & Negative & & Positive & & Negative \\
			\hline\hline
			\cellcolor{green!20} $+$ & \cellcolor{green!20}HillaryClinton & \cellcolor{blue!20} $\pm$ & \cellcolor{blue!20}Zika  & \cellcolor{green!20} $+$ & \cellcolor{green!20}America & \cellcolor{red!20} $-$ & \cellcolor{red!20}Obamacare  \\
			\cellcolor{green!20} $+$ & \cellcolor{green!20}POTUS & \cellcolor{red!20} $-$ & \cellcolor{red!20}Trump & \cellcolor{green!20} $+$ & \cellcolor{green!20}@SpeakerRyan & \cellcolor{blue!20} $\pm$ & \cellcolor{blue!20}\#BetterWay  \\
			\cellcolor{green!20} $+$ & \cellcolor{green!20}America &  \cellcolor{red!20} $-$ & \cellcolor{red!20}@HouseGOP & \cellcolor{green!20} $+$ & \cellcolor{green!20}Congress  & \cellcolor{blue!20} $\pm$ & \cellcolor{blue!20}Zika  \\
			\cellcolor{green!20} $+$ & \cellcolor{green!20}\#WomensEqualityDay & \cellcolor{red!20} $-$ & \cellcolor{red!20}Donald Trump  & \cellcolor{green!20} $+$ & \cellcolor{green!20}@Mike\_Pence  & \cellcolor{red!20} $-$ & \cellcolor{red!20}Iran \\
			\cellcolor{green!20} $+$ & \cellcolor{green!20}\#NationalComingOutDay & \cellcolor{red!20} $-$ & \cellcolor{red!20}Gun Violence  & \cellcolor{green!20} $+$ & \cellcolor{green!20}@RepTomPrice  &  \cellcolor{red!20} $-$ & \cellcolor{red!20}Obama \\
			\cellcolor{green!20} $+$ & \cellcolor{green!20}Americans & \cellcolor{blue!20} $\pm$ & \cellcolor{blue!20}\#Trans &  \cellcolor{green!20} $+$ & \cellcolor{green!20}@realDonaldTrump &  \cellcolor{red!20} $-$ & \cellcolor{red!20}Tax code  \\
			\cellcolor{green!20} $+$ & \cellcolor{green!20}\#LaborDay  & \cellcolor{red!20} $-$ & \cellcolor{red!20}\#GunViolence & \cellcolor{green!20} $+$ & \cellcolor{green!20}Texas  & \cellcolor{blue!20} $\pm$ & \cellcolor{blue!20}Breast Cancer  \\
			\cellcolor{green!20} $+$ & \cellcolor{green!20}TimKaine  & \cellcolor{blue!20} $\pm$ & \cellcolor{blue!20}Climate Change  & \cellcolor{green!20} $+$ & \cellcolor{green!20}\#VeteransDay  &  \cellcolor{red!20} $-$ & \cellcolor{red!20}President Obama \\
			\cellcolor{green!20} $+$ & \cellcolor{green!20}Hillary  &  \cellcolor{blue!20} $\pm$ & \cellcolor{blue!20}\#Trabajadores  & & ICYMI & \cellcolor{blue!20} $\pm$ & \cellcolor{blue!20}GITMO  \\
			\cellcolor{green!20} $+$ & \cellcolor{green!20}American & \cellcolor{blue!20} $\pm$ & \cellcolor{blue!20}TerenceCrutcher & & Senator &  \cellcolor{red!20} $-$ & \cellcolor{red!20}Islamic  \\
			& Cubs  & \cellcolor{red!20} $-$ & \cellcolor{red!20}GOP  &  \cellcolor{green!20} $+$ & \cellcolor{green!20}\#LaborDay & \cellcolor{red!20} $-$ & \cellcolor{red!20}State Sponsor \\
			\cellcolor{green!20} $+$ & \cellcolor{green!20}Halloween   & \cellcolor{red!20} $-$ & \cellcolor{red!20}Violence Situations &  \cellcolor{green!20} $+$ & \cellcolor{green!20}God &  \cellcolor{red!20} $-$ & \cellcolor{red!20}POTUS \\
			\cellcolor{green!20} $+$ & \cellcolor{green!20}Veterans & \cellcolor{red!20} $-$ & \cellcolor{red!20}ISIS & \cellcolor{green!20} $+$ & \cellcolor{green!20}Constitution Day &  \cellcolor{red!20} $-$ & \cellcolor{red!20}ISIS  \\
			& Florida  & \cellcolor{blue!20} $\pm$ & \cellcolor{blue!20}\#FundFlint &  \cellcolor{green!20} $+$ & \cellcolor{green!20}USMC& \cellcolor{red!20} $-$ & \cellcolor{red!20}Hillary   \\
			\cellcolor{green!20} $+$ & \cellcolor{green!20}\#LGBTQ equality &  \cellcolor{red!20} $-$ & \cellcolor{red!20}Donald & \cellcolor{green!20} $+$ & \cellcolor{green!20}Thanksgiving & \cellcolor{red!20} $-$ & \cellcolor{red!20}Fidel Castro \\
			\hline
		\end{tabular}
		\begin{tablenotes}
			\tiny
			\centering
			\item Note: All colors, signs, and the name of parties in the table are ground truth.
		\end{tablenotes}
	\end{threeparttable}
}
\end{table}

Table \ref{tab:US-motives} shows the profiles of two communities detected by our framework in the US dataset as well as their corresponding ground truth political parties and experts' labels. According to the provided background, the community on the left highly resembles the Democratic Party since its members have generally expressed: (1) positive attitudes towards Hillary Clinton, the U.S. president (i.e., POTUS), Tim Kaine, and human rights issues (e.g., \#WomensEqualityDay, \#LGBTQ equality, and \#NationalComingOutDay), and (2) negative attitudes towards the Republican Party (e.g., @HouseGOP and GOP), Donald Trump, and gun violence,  police brutality (e.g., the shooting of Terence Crutcher). On the other hand, the community on the right highly resembles the Republican Party since its members have generally expressed: (1) positive attitudes towards the Republican Party (e.g., @HouseGOP and @SpeakerRyan), Donald Trump, Mike Pence, Congress, and religion (i.e., God), and (2) negative attitudes towards President Obama and his policies (i.e., Obamacare, tax code, Iran, Guantanamo Bay detention camp (i.e., GITMO)) as well as Hillary Clinton.

\begin{table}[!htb]
	\centering
	\tiny
	\caption{The profiles of two communities detected by GNMF and DNMF in the US dataset.}
	\label{tab:US-profiles_baselines}
	\begin{minipage}{.5\linewidth}
		\centering
		\begin{threeparttable}
			\begin{tablenotes}
				\large
				\centering
				\item a. GNMF
				\tiny
				\linebreak
			\end{tablenotes}
			\setlength\tabcolsep{1.5pt}
			\begin{tabular}{| c c || c c |}
				\hline
				\multicolumn{2}{|c||}{Democrats} & \multicolumn{2}{c|}{Republicans} \\
				\hline\hline
				\cellcolor{red!20} $-$ & \cellcolor{red!20} Trump & \cellcolor{red!20} $-$ & \cellcolor{red!20} Obamacare \\
				\cellcolor{green!20} $+$ & \cellcolor{green!20}Hillary  & \cellcolor{blue!20} $\pm$ & \cellcolor{blue!20}\#BetterWay \\
				\cellcolor{green!20} $+$ & \cellcolor{green!20}Gov & \cellcolor{green!20} $+$ & \cellcolor{green!20}Congress \\
				\cellcolor{green!20} $+$ & \cellcolor{green!20}HillaryClinton & \cellcolor{blue!20} $\pm$ & \cellcolor{blue!20}Zika \\
				\cellcolor{red!20} $-$ & \cellcolor{red!20} Donald Trump & \cellcolor{green!20} $+$ & \cellcolor{green!20}America \\
				\cellcolor{blue!20} $\pm$ & \cellcolor{blue!20}\#DoYourJob & \cellcolor{green!20} $+$ & \cellcolor{green!20}@HouseGOP  \\
				& DebateNight &  \cellcolor{green!20} $+$ & \cellcolor{green!20}American   \\
				\cellcolor{red!20} $-$ & \cellcolor{red!20} @realDonaldTrump  & & ICYMI \\
				& China & \cellcolor{red!20} $-$ & \cellcolor{red!20} Obama  \\
				\cellcolor{red!20} $-$ & \cellcolor{red!20} @CoryBooker  &  & Florida \\
				\cellcolor{blue!20} $\pm$ & \cellcolor{blue!20}Russia & \cellcolor{green!20} $+$ & \cellcolor{green!20} Americans \\
				& ElectionDay &  \cellcolor{red!20} $-$ & \cellcolor{red!20} Iran \\
				\cellcolor{blue!20} $\pm$ & \cellcolor{blue!20}Climate Change  & \cellcolor{green!20} $+$ & \cellcolor{green!20}U.S. \\
				& Debate &  \cellcolor{blue!20} $\pm$ & \cellcolor{blue!20}\#HurricanMatthew \\
				\cellcolor{green!20} $+$ & \cellcolor{green!20}Hillary Clinton  & \cellcolor{red!20} $-$ & \cellcolor{red!20} POTUS \\
				\cellcolor{red!20} $-$ & \cellcolor{red!20} Donald  & \cellcolor{green!20} $+$ & \cellcolor{green!20}@realDonaldTrump \\
				& Virginia  & \cellcolor{red!20} $-$ & \cellcolor{red!20} Clinton \\
				\cellcolor{green!20} $+$ & \cellcolor{green!20}HRC  & \cellcolor{blue!20} $\pm$ & \cellcolor{blue!20}Hurrican Matthew \\
				& VPDebate & \cellcolor{red!20} $-$ & \cellcolor{red!20}  Washington \\
				\cellcolor{green!20} $+$ & \cellcolor{green!20}America & \cellcolor{blue!20} $\pm$ & \cellcolor{blue!20}FBI \\
				\cellcolor{green!20} $+$ & \cellcolor{green!20}TimKaine  &  \cellcolor{green!20} $+$ & \cellcolor{green!20}Texas  \\
				\cellcolor{green!20} $+$ & \cellcolor{green!20}\#WomenEqualityDay  &  \cellcolor{green!20} $+$ & \cellcolor{green!20}Senate \\
				\cellcolor{green!20} $+$ & \cellcolor{green!20}FLOTUS   & \cellcolor{green!20} $+$ & \cellcolor{green!20}Veterans \\
				\cellcolor{green!20} $+$ & \cellcolor{green!20}\#IamWithHer &  \cellcolor{blue!20} $\pm$ & \cellcolor{blue!20}Matthew \\
				\cellcolor{blue!20} $\pm$ & \cellcolor{blue!20}Flint & & \#DoYourJob\\
				\cellcolor{green!20} $+$ & \cellcolor{green!20}POTUS & \cellcolor{green!20} $+$ & \cellcolor{green!20}@SpeakerRyan \\
				\cellcolor{green!20} $+$ & \cellcolor{green!20}HouseDemocrats  & & Ohio  \\
				\cellcolor{red!20} $-$ & \cellcolor{red!20}  Steve Bannon & \cellcolor{blue!20} $\pm$ & \cellcolor{blue!20}\#NeverForget \\
				\cellcolor{red!20} $-$ & \cellcolor{red!20} Bannon &  \cellcolor{green!20} $+$ & \cellcolor{green!20}GOP \\
				\cellcolor{green!20} $+$ & \cellcolor{green!20}USA & & WSJ \\
				\hline
			\end{tabular}
		\end{threeparttable}
	\end{minipage}%
	\begin{minipage}{.5\linewidth}
		\centering
		\begin{threeparttable}
			\begin{tablenotes}
				\large
				\centering
				\item b. DNMF
				\tiny
				\linebreak
			\end{tablenotes}
			\setlength\tabcolsep{1.5pt}
			\begin{tabular}{| c c || c c |}
				\hline
				\multicolumn{2}{|c||}{Democrats} & \multicolumn{2}{c|}{Republicans} \\
				\hline\hline
				\cellcolor{red!20} $-$ & \cellcolor{red!20} Congress &  \cellcolor{green!20} $+$ & \cellcolor{green!20}Congress\\
				\cellcolor{green!20} $+$ & \cellcolor{green!20}Obamacare &  \cellcolor{red!20} $-$ & \cellcolor{red!20} Obamacare \\
				\cellcolor{red!20} $-$ & \cellcolor{red!20} Trump &  \cellcolor{green!20} $+$ & \cellcolor{green!20}Trump \\
				& \#BetterWay &  \cellcolor{blue!20} $\pm$ & \cellcolor{blue!20}\#BetterWay \\
				\cellcolor{green!20} $+$ & \cellcolor{green!20}America &  \cellcolor{green!20} $+$ & \cellcolor{green!20}America \\
				\cellcolor{blue!20} $\pm$ & \cellcolor{blue!20}Zika &  \cellcolor{blue!20} $\pm$ & \cellcolor{blue!20}Zika \\
				\cellcolor{red!20} $-$ & \cellcolor{red!20} @HouseGOP &  \cellcolor{green!20} $+$ & \cellcolor{green!20}@HouseGOP \\
				\cellcolor{green!20} $+$ & \cellcolor{green!20}American &  \cellcolor{green!20} $+$ & \cellcolor{green!20}American \\
				\cellcolor{green!20} $+$ & \cellcolor{green!20}Gov &  \cellcolor{red!20} $-$ & \cellcolor{red!20} Gov \\
				\cellcolor{blue!20} $\pm$ & \cellcolor{blue!20}\#DoYourJob &  & \#DoYourJob \\
				& ICYMI & & ICYMI \\
				\cellcolor{green!20} $+$ & \cellcolor{green!20}Americans &  \cellcolor{red!20} $-$ & \cellcolor{red!20} HillaryClinton \\
				\cellcolor{green!20} $+$ & \cellcolor{green!20}HillaryClinton & \cellcolor{green!20} $+$ & \cellcolor{green!20}Americans \\
				\cellcolor{green!20} $+$ & \cellcolor{green!20}Hillary &  \cellcolor{red!20} $-$ & \cellcolor{red!20} Hillary \\
				\cellcolor{red!20} $-$ & \cellcolor{red!20} @realDonaldTrump &  \cellcolor{green!20} $+$ & \cellcolor{green!20}@realDonaldTrump \\
				\cellcolor{green!20} $+$ & \cellcolor{green!20}Obama &  \cellcolor{red!20} $-$ & \cellcolor{red!20} Obama \\
				\cellcolor{green!20} $+$ & \cellcolor{green!20}U.S. &  \cellcolor{green!20} $+$ & \cellcolor{green!20}U.S. \\
				\cellcolor{green!20} $+$ & \cellcolor{green!20}POTUS &  \cellcolor{red!20} $-$ & \cellcolor{red!20} POTUS \\
				\cellcolor{green!20} $+$ & \cellcolor{green!20}Iran &  \cellcolor{red!20} $-$ & \cellcolor{red!20} Iran \\
				\cellcolor{green!20} $+$ & \cellcolor{green!20}Clinton &  \cellcolor{red!20} $-$ & \cellcolor{red!20} Clinton \\
				\cellcolor{red!20} $-$ & \cellcolor{red!20} Donald Trump &  \cellcolor{green!20} $+$ & \cellcolor{green!20}Donald Trump \\
				\cellcolor{green!20} $+$ & \cellcolor{green!20}Veterans &  \cellcolor{green!20} $+$ & \cellcolor{green!20}Veterans \\
				\cellcolor{green!20} $+$ & \cellcolor{green!20}Washington &  \cellcolor{red!20} $-$ & \cellcolor{red!20} Washington \\
				\cellcolor{blue!20} $\pm$ & \cellcolor{blue!20}HurricanMatthew &  \cellcolor{blue!20} $\pm$ & \cellcolor{blue!20}HurricanMatthew \\
				\cellcolor{red!20} $-$ & \cellcolor{red!20} Senate &  \cellcolor{green!20} $+$ & \cellcolor{green!20}Senate \\
				\cellcolor{blue!20} $\pm$ & \cellcolor{blue!20}FBI &  \cellcolor{blue!20} $\pm$ & \cellcolor{blue!20}FBI \\
				& Florida & &  Florida \\
				& Texas &  \cellcolor{green!20} $+$ & \cellcolor{green!20}Texas \\
				\cellcolor{red!20} $-$ & \cellcolor{red!20} GOP &  \cellcolor{green!20} $+$ & \cellcolor{green!20}GOP \\
				& Oct & & Oct \\
				\hline
			\end{tabular}
		\end{threeparttable}
	\end{minipage} 
\end{table}

Negative sentiment implies both opposition and concern. If necessary, our framework can differentiate opposition from concern by providing the sentiment words frequently expressed by the members of a community towards each key expression. For example, 
Democrats' negative sentiment towards Donald Trump mainly comes from the sentiment words ``unfit'', ``low'', and ``dangerous'' which suggest opposition. On the other hand, their negative sentiment towards \#Trans (i.e., transgender people) mainly originates from the sentiment words ``discrimination'' and ``murder'' which indicate concern.


To demonstrate the advantage of our community profiling method, we compare the profiles of typical community profiles usually provided by retrospective studies with those uncovered by our framework. Table \ref{tab:US-profiles_baselines} shows the profiles of two communities detected by GNMF and DNMF in the US dataset as well as their corresponding ground truth political parties. As we observe, it is almost impossible for a non-expert individual to recognize the party associated with each profile since the position of the communities towards the key expressions are not taken into account. For example, in profiles corresponding to the Democratic Party and the Republican Party, many key expressions related to Trump, Clinton, and Obama exist, but there is no information regarding collective attitude of community members toward such key expressions. However, Table \ref{tab:US-motives} shows that our proposed method correctly divides opposed/concerned key expressions and supported ones into the correct categories. Therefore, our framework makes it easy not only to differentiate and understand communities better but also to associate online communities with their real-world counterparts (if exist).

\begin{table*}
	\centering
	\tiny
	\caption{The profiles of five communities detected by our framework in the UK dataset.}
	\label{tab:UK-motives}
	\setlength\tabcolsep{1.5pt}
	\begin{tabular}{| cc | cc || cc | cc || cc | cc || cc | cc || cc | cc |}
		
		\hline
		
		\multicolumn{4}{|c||}{Conservative Party (Tory)} & \multicolumn{4}{|c||}{Labour Party (Lab)}  & \multicolumn{4}{|c|}{Liberal Democrat Party (Lib Dem)} & \multicolumn{4}{|c||}{Scottish National Party (SNP)} &  \multicolumn{4}{|c|}{UK Independence Party (UKIP)}\\ 
		
		\hline
		
		& Positive & & Negative & & Positive & & Negative & & Positive & & Negative & & Positive & & Negative & & Positive & & Negative \\
		
		\hline\hline
		
		\cellcolor{green!20} $+$ & \cellcolor{green!20}Conservatives & \cellcolor{red!20} $-$ & \cellcolor{red!20} Labour & \cellcolor{green!20} $+$ & \cellcolor{green!20}Labour & \cellcolor{red!20} $-$ & \cellcolor{red!20} Tories & \cellcolor{green!20} $+$ & \cellcolor{green!20}LibDems & \cellcolor{red!20} $-$ & \cellcolor{red!20} Labour & \cellcolor{green!20} $+$ & \cellcolor{green!20}TheSNP & \cellcolor{red!20} $-$ & \cellcolor{red!20} Tory  & \cellcolor{green!20} $+$ & \cellcolor{green!20}UKIP & \cellcolor{blue!20} $\pm$ & \cellcolor{blue!20}Calais     \\
		
		\cellcolor{green!20} $+$ & \cellcolor{green!20}@David\_Cameron & \cellcolor{red!20} $-$ & \cellcolor{red!20} Miliband  & \cellcolor{green!20} $+$ & \cellcolor{green!20}UKLabour & \cellcolor{red!20} $-$ & \cellcolor{red!20} Tory & \cellcolor{green!20} $+$ & \cellcolor{green!20}@Nick\_Clegg & \cellcolor{blue!20} $\pm$ & \cellcolor{blue!20}Iraq & & GE15 & \cellcolor{blue!20} $\pm$ & \cellcolor{blue!20}Trident & \cellcolor{green!20} $+$ & \cellcolor{green!20}@Nigel\_Farage' & \cellcolor{red!20} $-$ & \cellcolor{red!20} Labour \\
		
		\cellcolor{green!20} $+$ & \cellcolor{green!20}Cameron & \cellcolor{blue!20} $\pm$ & \cellcolor{blue!20}Tunisia & \cellcolor{green!20} $+$ & \cellcolor{green!20}LabourDoorStep & \cellcolor{red!20} $-$ & \cellcolor{red!20} Cameron & & GE2015 & \cellcolor{blue!20} $\pm$ & \cellcolor{blue!20}Climate Change & \cellcolor{green!20} $+$ & \cellcolor{green!20}SNP & \cellcolor{red!20} $-$ & \cellcolor{red!20} Tories & \cellcolor{green!20} $+$ & \cellcolor{green!20}Nigel Farage & \cellcolor{red!20} $-$ & \cellcolor{red!20} ISIS \\
		
		& GE2015 & \cellcolor{blue!20} $\pm$ & \cellcolor{blue!20}Paris & \cellcolor{green!20} $+$ & \cellcolor{green!20}@AndyBurnhamMP & & A\&E & \cellcolor{green!20} $+$ & \cellcolor{green!20}NormanLamb& \cellcolor{red!20} $-$ & \cellcolor{red!20} Tories & \cellcolor{green!20} $+$ & \cellcolor{green!20}NicolaSturgeon & \cellcolor{red!20} $-$ & \cellcolor{red!20} Labour & & BBCqt &  \cellcolor{blue!20} $\pm$ & \cellcolor{blue!20}Greece\\
		
		\cellcolor{green!20} $+$ & \cellcolor{green!20}@George\_Osborne & & FIFA & \cellcolor{green!20} $+$ & \cellcolor{green!20}Ed\_Miliband& \cellcolor{red!20} $-$ & \cellcolor{red!20} David Cameron & & Cardiff & \cellcolor{red!20} $-$ & \cellcolor{red!20} Tory & \cellcolor{green!20} $+$ & \cellcolor{green!20}Scotland & \cellcolor{green!20} $+$ & \cellcolor{green!20}\#RefugeesWelcome & & GE2015 & \cellcolor{red!20} $-$ & \cellcolor{red!20} BBC \\
		
		\cellcolor{green!20} $+$ & \cellcolor{green!20}VoteConservative$^\star$ &  \cellcolor{red!20} $-$ & \cellcolor{red!20} ISIL  & \cellcolor{green!20} $+$ & \cellcolor{green!20}Britain & \cellcolor{red!20} $-$ & \cellcolor{red!20} Bedroom Tax & \cellcolor{green!20} $+$ & \cellcolor{green!20}@TimFarron & \cellcolor{red!20} $-$ & \cellcolor{red!20} UKLabour & & MPs & \cellcolor{blue!20} $\pm$ & \cellcolor{blue!20}Iraq & \cellcolor{red!20} $-$ & \cellcolor{red!20} Cameron & \cellcolor{blue!20} $\pm$ & \cellcolor{blue!20}Britain \\
		
		\cellcolor{green!20} $+$ & \cellcolor{green!20}NickyMorgan01$^\star$ & \cellcolor{blue!20} $\pm$ & \cellcolor{blue!20}Heathrow & \cellcolor{green!20} $+$ & \cellcolor{green!20}@GloriadePiero & \cellcolor{red!20} $-$ & \cellcolor{red!20} BedroomTax & \cellcolor{green!20} $+$ & \cellcolor{green!20}Lib Dem & \cellcolor{red!20} $-$ & \cellcolor{red!20} Tuition Fees & \cellcolor{green!20} $+$ & \cellcolor{green!20}Glasgow & \cellcolor{blue!20} $\pm$ & \cellcolor{blue!20}Paris & & Mark& \cellcolor{red!20} $-$ & \cellcolor{red!20} Government\\
		
		& London & \cellcolor{red!20} $-$ & \cellcolor{red!20} Charles Kennedy & \cellcolor{green!20} $+$ & \cellcolor{green!20}YvetteCooperMP& \cellcolor{red!20} $-$ & \cellcolor{red!20} Govt & \cellcolor{blue!20} $\pm$ & \cellcolor{blue!20}Mental Health & \cellcolor{blue!20} $\pm$ & \cellcolor{blue!20}HIV & & Alan & \cellcolor{blue!20} $\pm$ & \cellcolor{blue!20}CharlieHebdo$^\star$ & \cellcolor{green!20} $+$ & \cellcolor{green!20}Brexit  & \cellcolor{red!20} $-$ & \cellcolor{red!20} Tories \\
		
		& Wales &  \cellcolor{red!20} $-$ & \cellcolor{red!20} Ed\_Miliband& \cellcolor{green!20} $+$ & \cellcolor{green!20}YvetteForLabour$^\star$  & \cellcolor{blue!20} $\pm$ & \cellcolor{blue!20}Tax Credits &  \cellcolor{green!20} $+$ & \cellcolor{green!20}Lib Dems & \cellcolor{red!20} $-$ & \cellcolor{red!20} SNP & \cellcolor{green!20} $+$ & \cellcolor{green!20}Scottish & \cellcolor{blue!20} $\pm$ & \cellcolor{blue!20}Syria & & Telegraph &  \cellcolor{red!20} $-$ & \cellcolor{red!20} David Cameron\\
		
		\cellcolor{green!20} $+$ & \cellcolor{green!20}David Cameron &  \cellcolor{red!20} $-$ & \cellcolor{red!20} SNP   & \cellcolor{green!20} $+$ & \cellcolor{green!20}SteveReedMP & \cellcolor{red!20} $-$ & \cellcolor{red!20} SNP & \cellcolor{green!20} $+$ & \cellcolor{green!20}NHS & &  PMQs & & GE2015 & \cellcolor{blue!20} $\pm$ & \cellcolor{blue!20}Mediterranean & & BBC5live& \cellcolor{blue!20} $\pm$ & \cellcolor{blue!20}Libya \\
		
		\cellcolor{green!20} $+$ & \cellcolor{green!20}@Jeremy\_Hunt &  \cellcolor{blue!20} $\pm$ & \cellcolor{blue!20}Syria  & \cellcolor{green!20} $+$ & \cellcolor{green!20}@LeicesterLiz & \cellcolor{red!20} $-$ & \cellcolor{red!20} Government &  & John& \cellcolor{blue!20} $\pm$ & \cellcolor{blue!20}CharlieHebdo & & Edinburgh  & \cellcolor{blue!20} $\pm$ & \cellcolor{blue!20}Med & \cellcolor{green!20} $+$ & \cellcolor{green!20}Queen &  \cellcolor{red!20} $-$ & \cellcolor{red!20} Miliband \\
		
		\cellcolor{green!20} $+$ & \cellcolor{green!20}Team2015$^\star$ & \cellcolor{red!20} $-$ & \cellcolor{red!20} Lab  &  \cellcolor{green!20} $+$ & \cellcolor{green!20}LizforLeader$^\star$ & \cellcolor{blue!20} $\pm$ & \cellcolor{blue!20}France & \cellcolor{green!20} $+$ & \cellcolor{green!20}Nick Clegg & \cellcolor{red!20} $-$ & \cellcolor{red!20} Nigel Farage & & Maiden Speech &  \cellcolor{blue!20} $\pm$ & \cellcolor{blue!20}French & & Andrew&  \cellcolor{blue!20} $\pm$ & \cellcolor{blue!20}Tunisia \\
		
		& Chris & \cellcolor{red!20} $-$ & \cellcolor{red!20} LibDems & \cellcolor{green!20} $+$ & \cellcolor{green!20}@TristramHuntMP & \cellcolor{blue!20} $\pm$ & \cellcolor{blue!20}Tunisia & & LBC& \cellcolor{blue!20} $\pm$ & \cellcolor{blue!20}Ebola & \cellcolor{green!20} $+$ & \cellcolor{green!20}NHS & \cellcolor{blue!20} $\pm$ & \cellcolor{blue!20}Charles & \cellcolor{green!20} $+$ & \cellcolor{green!20}George's  & \cellcolor{blue!20} $\pm$ & \cellcolor{blue!20}Paris \\
		
		\cellcolor{green!20} $+$ & \cellcolor{green!20}@Tracey\_Crouch & \cellcolor{blue!20} $\pm$ & \cellcolor{blue!20}Calais & & GE2015 & \cellcolor{blue!20} $\pm$ & \cellcolor{blue!20}Syria & \cellcolor{green!20} $+$ & \cellcolor{green!20}LibDem & \cellcolor{blue!20} $\pm$ & \cellcolor{blue!20}Paris & & Neil& \cellcolor{blue!20} $\pm$ & \cellcolor{blue!20}Tunisia & \cellcolor{red!20} $-$ & \cellcolor{red!20} JeremyCorbyn$^\star$ & \cellcolor{red!20} $-$ & \cellcolor{red!20} SNP \\
		
		\cellcolor{green!20} $+$ & \cellcolor{green!20}England & \cellcolor{blue!20} $\pm$ & \cellcolor{blue!20}Nepal & \cellcolor{green!20} $+$ & \cellcolor{green!20}VoteLabour$^\star$ & \cellcolor{green!20} $+$ & \cellcolor{green!20}Europe & \cellcolor{green!20} $+$ & \cellcolor{green!20}Govt & \cellcolor{red!20} $-$ & \cellcolor{red!20}  Welfare Bill & \cellcolor{green!20} $+$ & \cellcolor{green!20}Nicola Sturgeon & \cellcolor{red!20} $-$ & \cellcolor{red!20} Westminster & \cellcolor{red!20} $-$ & \cellcolor{red!20} Jeremy Corbyn & \cellcolor{blue!20} $\pm$ & \cellcolor{blue!20}Mediterranean\\
		
		\hline
		
		\noalign{\smallskip}
		
	\end{tabular}

		Note: All colors, signs, and the name of parties in the table are ground truth.	
\end{table*}

\subsubsection{UK Politics}
\label{UK_Politics}
The UK dataset covers many events such as the rise of terrorism and terrorist attacks (e.g., CharlieHebdo and Tunisia attack), and many natural disasters (e.g., Nepal earthquake and Ebola) that happened in the first nine months of 2015. However, the major event in this period of time is the UK general election. Brief background knowledge about five major UK parties during the general election are provided as follows \cite{moran2015politics},
\begin{itemize}
	\item \textbf{Conservative Party:} This party is also known as Tory and was led by David Cameron in 2015. David Cameron also led the UK government before and after the election of 2015. George Osborne, Nicky Morgan, and Jeremy Hunt were some of his secretaries.
	\item \textbf{Labour Party:} Ed Miliband was the leader of the Labour party for the election and selected Tom Watson as his deputy chair and campaign coordinator. Jeremy Corbyn, Yvette Cooper, Liz Kendall, and Andy Burnham were among the prominent members of the party.
	\item \textbf{Liberal Democrat Party:} Nick Clegg led the Liberal Democrat Party in 2015. Norman Lamb, John Leech, Nick Harvey, Tim Farron, and Charles Kennedy were some of the party's parliamentarians.
	\item \textbf{Scottish National Party:} The SNP is a Scottish Nationalist party led by Nicola Sturgeon in 2015. Alan Brown and Neil Gray were some of the party's parliamentarians.
	
	\item \textbf{UK Independence Party:} UKIP was led by Nigel Farage in 2015. The  party embodies opposition to both United Kingdom EU membership and immigration.
\end{itemize}

Table \ref{tab:UK-motives} shows the profiles of five communities detected by our framework in the UK dataset as well as their corresponding ground truth political parties. As shown in the table, all parties have a common key expression, the general election of 2015 (e.g, GE2015 and GE15). We can also observe that the members of each party have generally expressed: (1) positive attitudes towards their party and also their prominent members and (2) negative attitudes towards other parties and their prominent members due to election competition \cite{moran2015politics}. Moreover, the government was a coalition between the Conservative Party and the Liberal Democrat Party before the election. This coalition explains why they expressed positive sentiments towards the government related issues (i.e., Govt and Cameron) \cite{moran2015politics}. According to the negative attitudes of almost all parties, we can determine that the Conservative party and the Labour party are the ones towards which other parties expressed most negative sentiments. Furthermore, these two parties expressed a high negative sentiment towards each other. The reason behind this antagonism is that these parties are the two biggest parties having the highest chance of winning an outright majority in the election \cite{moran2015politics}. Moreover, UKIP's negative view on Calais, the city in France where immigrants enter the UK, and Mediterranean (immigrants/immigration) reflects its anti-immigration stance. Moreover, UKIP's positive sentiment on Brexit and its negative sentiment on Greece indicates its anti-EU orientation.

\begin{table}
	\centering
	\caption{The profiles of five communities detected by using GNMF in the UK dataset .}
	\label{tab:UK-GNMF}
	\setlength\tabcolsep{1.5pt}
	\resizebox{\columnwidth}{!}{%
		\begin{tabular}{| cc | cc | cc | cc | cc | cc | cc | cc | cc | cc |}
			\hline
			\multicolumn{2}{|c|}{Conservatives} & \multicolumn{2}{|c|}{Labours}  & \multicolumn{2}{|c|}{Lib dems} & \multicolumn{2}{|c|}{SNPs} &  \multicolumn{2}{|c|}{UKIPs}\\
			
			\hline\hline
			
			\cellcolor{red!20} $-$ & \cellcolor{red!20} Miliband & \cellcolor{red!20} $-$ & \cellcolor{red!20} UKIP & \cellcolor{green!20} $+$ & \cellcolor{green!20}Libdems & \cellcolor{green!20} $+$ & \cellcolor{green!20}SNP &\cellcolor{red!20} $-$ & \cellcolor{red!20} @JessPhillips \\ 
			\cellcolor{green!20} $+$ & \cellcolor{green!20}Conservatives & \cellcolor{green!20} $+$ & \cellcolor{green!20}Labour & & GE2015 & \cellcolor{green!20} $+$ & \cellcolor{green!20}Scotland & & birmingham \\ 
			& GE2015 & \cellcolor{red!20} $-$ & \cellcolor{red!20} Tories & \cellcolor{red!20} $-$ & \cellcolor{red!20} Labour& \cellcolor{green!20} $+$ & \cellcolor{green!20}VoteSNP  & \cellcolor{red!20} $-$ & \cellcolor{red!20} Labour \\
			\cellcolor{green!20} $+$ & \cellcolor{green!20}@David\_Cameron &  \cellcolor{green!20} $+$ & \cellcolor{green!20}NHS & \cellcolor{green!20} $+$ & \cellcolor{green!20}@LFeatherstone & & GE15 & & john \\
			\cellcolor{red!20} $-$ & \cellcolor{red!20} Labour &  \cellcolor{green!20} $+$ & \cellcolor{green!20}Britain & \cellcolor{red!20} $-$ & \cellcolor{red!20} @CLeslieMP & \cellcolor{green!20} $+$ & \cellcolor{green!20}@TheSNP & \cellcolor{red!20} $-$ & \cellcolor{red!20} Libdems \\
			\cellcolor{green!20} $+$ & \cellcolor{green!20}VoteConservetives &  \cellcolor{red!20} $-$ & \cellcolor{red!20} Cameron & & Bradford & \cellcolor{red!20} $-$ & \cellcolor{red!20} Labour & \cellcolor{red!20} $-$ & \cellcolor{red!20} NHS \\
			\cellcolor{green!20} $+$ & \cellcolor{green!20}Govt &  \cellcolor{red!20} $-$ & \cellcolor{red!20} @Nigel\_Frange & \cellcolor{green!20} $+$ & \cellcolor{green!20}@Nick\_Clegg & \cellcolor{red!20} $-$ & \cellcolor{red!20} Westminster & \cellcolor{red!20} $-$ & \cellcolor{red!20} LabourEoin \\
			\cellcolor{green!20} $+$ & \cellcolor{green!20}NHS &  \cellcolor{green!20} $+$ & \cellcolor{green!20}UKLabour & & London & \cellcolor{red!20} $-$ & \cellcolor{red!20} Tory & \cellcolor{red!20} $-$ & \cellcolor{red!20} Lib dems \\
			& LeaderDebates & &  London & \cellcolor{blue!20} $\pm$ & \cellcolor{blue!20}Budget2015 & & GE2015 & \cellcolor{red!20} $-$ & \cellcolor{red!20} Lib dem \\
			\cellcolor{green!20} $+$ & \cellcolor{green!20}@ZacGoldsmith &  \cellcolor{green!20} $+$ & \cellcolor{green!20}LabourDoorStep & & Wales & \cellcolor{green!20} $+$ & \cellcolor{green!20}@NicolaSturgeon & & Hansard \\
			\cellcolor{red!20} $-$ & \cellcolor{red!20} @Ed\_Miliband & & BBC & \cellcolor{green!20} $+$ & \cellcolor{green!20}NHS & & LeadersDebate & \cellcolor{red!20} $-$ & \cellcolor{red!20} @TobyPerkinsMP \\
			& MPs & & BBCqt & \cellcolor{red!20} $-$ & \cellcolor{red!20} Miliband & \cellcolor{red!20} $-$ & \cellcolor{red!20} LaboursDoorStep & \cellcolor{red!20} $-$ & \cellcolor{red!20} UKLabour \\
			& London &  \cellcolor{green!20} $+$ & \cellcolor{green!20}Europe & & LeaderDebate & & MPs & & MPs \\
			\cellcolor{red!20} $-$ & \cellcolor{red!20} UKLabour &  \cellcolor{green!20} $+$ & \cellcolor{green!20}@AndyBurnhamMP & \cellcolor{red!20} $-$ & \cellcolor{red!20} @George\_Osborne & \cellcolor{blue!20} $\pm$ & \cellcolor{blue!20}Trident & & @SabelHardman \\
			& Wales &  \cellcolor{green!20} $+$ & \cellcolor{green!20}@ED\_Miliband & \cellcolor{green!20} $+$ & \cellcolor{green!20}Lib dems & \cellcolor{green!20} $+$ & \cellcolor{green!20}Scottish & & Jess \\
			\cellcolor{green!20} $+$ & \cellcolor{green!20}England &  \cellcolor{green!20} $+$ & \cellcolor{green!20}TessaJowell & \cellcolor{red!20} $-$ & \cellcolor{red!20} David Cameron & & London & \cellcolor{red!20} $-$ & \cellcolor{red!20} Labour party \\
			\cellcolor{green!20} $+$ & \cellcolor{green!20}@George\_Osborne &  \cellcolor{red!20} $-$ & \cellcolor{red!20} David Cameron & \cellcolor{green!20} $+$ & \cellcolor{green!20}Lib dem & \cellcolor{red!20} $-$ & \cellcolor{red!20} UKLabour & \cellcolor{red!20} $-$ & \cellcolor{red!20} @SimonDanczuk \\
			& croydon & \cellcolor{red!20} $-$ & \cellcolor{red!20} Tory &  \cellcolor{blue!20} $\pm$ & \cellcolor{blue!20}Mental Health & & PMQS & \cellcolor{red!20} $-$ & \cellcolor{red!20} Libdem \\
			\cellcolor{green!20} $+$ & \cellcolor{green!20}@NickyMorgan01 &  \cellcolor{green!20} $+$ & \cellcolor{green!20}Corbyn & \cellcolor{green!20} $+$ & \cellcolor{green!20}@SWilliamsMP & \cellcolor{red!20} $-$ & \cellcolor{red!20} @David\_Cameron & & GE2015 \\
			\cellcolor{green!20} $+$ & \cellcolor{green!20}@NorwichChloe &  \cellcolor{blue!20} $\pm$ & \cellcolor{blue!20}Calasis & \cellcolor{green!20} $+$ & \cellcolor{green!20}@NormanLamb & & Wales & & Youtube \\
			\cellcolor{green!20} $+$ & \cellcolor{green!20}@RobertBuckland &  \cellcolor{green!20} $+$ & \cellcolor{green!20}@YvetteCooper & \cellcolor{red!20} $-$ & \cellcolor{red!20} Conservatives & \cellcolor{green!20} $+$ & \cellcolor{green!20}@GradySNP & \cellcolor{red!20} $-$ & \cellcolor{red!20} Europe \\
			\cellcolor{red!20} $-$ & \cellcolor{red!20} VoteLabour &  \cellcolor{green!20} $+$ & \cellcolor{green!20}VoteLabour & \cellcolor{red!20} $-$ & \cellcolor{red!20} Tories & \cellcolor{green!20} $+$ & \cellcolor{green!20}Glasgow & \cellcolor{red!20} $-$ & \cellcolor{red!20} miliband \\
			\cellcolor{red!20} $-$ & \cellcolor{red!20} @CLeslieMP &  \cellcolor{blue!20} $\pm$ & \cellcolor{blue!20}Greece & \cellcolor{green!20} $+$ & \cellcolor{green!20}Nick Clegg & & Front Page & \cellcolor{red!20} $-$ & \cellcolor{red!20} Food Banks \\
			\cellcolor{red!20} $-$ & \cellcolor{red!20} LabourLeadership &  \cellcolor{green!20} $+$ & \cellcolor{green!20}England & & Cardiff & \cellcolor{red!20} $-$ & \cellcolor{red!20} VoteLabour & \cellcolor{red!20} $-$ & \cellcolor{red!20}  Housing Benefit \\
			\cellcolor{green!20} $+$ & \cellcolor{green!20}Tories &  \cellcolor{green!20} $+$ & \cellcolor{green!20}Jeremy Corbyn & \cellcolor{green!20} $+$ & \cellcolor{green!20}@TimFarron & \cellcolor{red!20} $-$ & \cellcolor{red!20} ScottishLabour & & Google \\
			\cellcolor{green!20} $+$ & \cellcolor{green!20}Minister &  \cellcolor{red!20} $-$ & \cellcolor{red!20} Farage & \cellcolor{red!20} $-$ & \cellcolor{red!20} VoteConservatives & \cellcolor{green!20} $+$ & \cellcolor{green!20}Nicola Sturgeon & \cellcolor{red!20} $-$ & \cellcolor{red!20} @GiselaStuart \\
			\cellcolor{red!20} $-$ & \cellcolor{red!20} Guardian &  \cellcolor{green!20} $+$ & \cellcolor{green!20}YvetteCooperLabour & & Bristol & \cellcolor{red!20} $-$ & \cellcolor{red!20} LabourLeadership & \cellcolor{red!20} $-$ & \cellcolor{red!20} Labour MPs \\
			& Leeds & \cellcolor{red!20} $-$ & \cellcolor{red!20} Telegraph &  & Croydon & \cellcolor{red!20} $-$ & \cellcolor{red!20} Lab & \cellcolor{green!20} $+$ & \cellcolor{green!20}Britain \\
			\cellcolor{green!20} $+$ & \cellcolor{green!20}Government &  \cellcolor{green!20} $+$ & \cellcolor{green!20}@EmmaReynoldMP & & Norwich & & Edinburgh & & Wales \\
			& State & & Thurrock &  \cellcolor{red!20} $-$ & \cellcolor{red!20} Chancellor & \cellcolor{red!20} $-$ & \cellcolor{red!20} @AndyburnhamMP & \cellcolor{red!20} $-$ & \cellcolor{red!20} @David\_Cameron \\

			\hline
			
			\noalign{\smallskip}
		\end{tabular}
	}	
	\tiny
	Note: All colors, signs, and the name of parties in the table are ground truth.
	
\end{table}


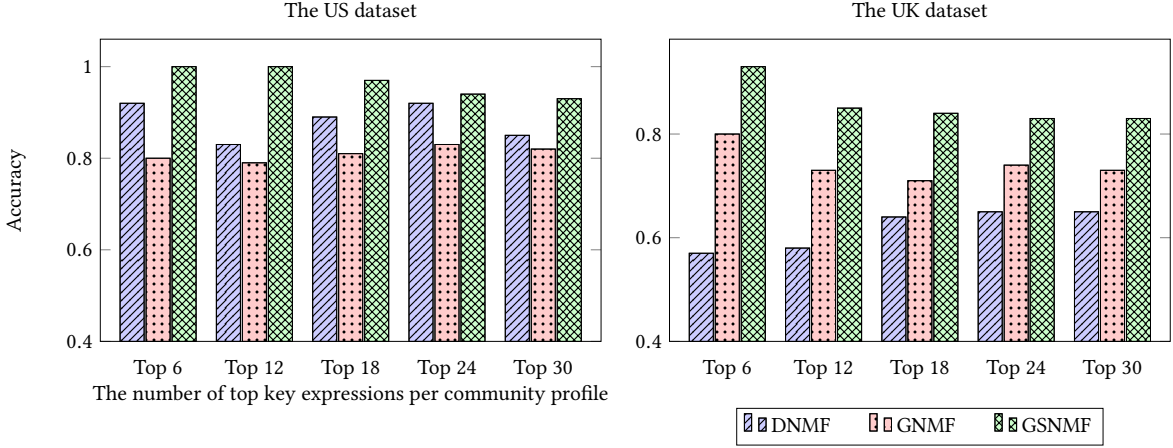
\begin{figure*}
	\centering
	\scalebox{0.91}{
		\begin{tikzpicture}
		\begin{groupplot}[group style={group size= 2 by 4},
		width  = 0.5*\textwidth,
		height = 6cm,
		major x tick style = transparent,
		ybar=2*\pgflinewidth,
		symbolic x coords={Top 6, Top 12, Top 18, Top 24, Top 30},
		xtick = data,
		scaled y ticks = false,
		enlarge x limits=0.15,
		ymin=0.4,
		legend cell align=left,
		legend style={
			at={(0.5,-0.22)},
			anchor=north,
			legend columns=-1,
			/tikz/every even column/.append style={column sep=0.5cm}
		},
		]
		
		\nextgroupplot[title=The US dataset,ylabel={Accuracy}, 
		xlabel = {The number of top key expressions per community profile}]
		
		\addplot +[
		black,
		fill=blue!20,
		postaction={
			pattern=north east lines
		}
		]
		coordinates {(Top 6, 0.92) (Top 12, 0.83) (Top 18, 0.89) (Top 24, 0.92) (Top 30, 0.85)};
		
		\addplot +[
		black,
		fill=red!20,
		postaction={
			pattern=dots
		}
		]
		coordinates {(Top 6, 0.8) (Top 12, 0.79) (Top 18, 0.81) (Top 24, 0.83) (Top 30, 0.82)};
		
		\addplot +[
		black,
		fill=green!20,
		postaction={
			pattern=crosshatch
		}
		]
		coordinates {(Top 6, 1) (Top 12, 1) (Top 18, 0.97) (Top 24, 0.94) (Top 30, 0.93)};
		
		\nextgroupplot[title=The UK dataset]
		
		\addplot +[
		black,
		fill=blue!20,
		postaction={
			pattern=north east lines
		}
		]
		coordinates {(Top 6, 0.57) (Top 12, 0.58) (Top 18, 0.64) (Top 24, 0.65) (Top 30, 0.65)};
		
		\addplot +[
		black,
		fill=red!20,
		postaction={
			pattern=dots
		}
		]
		coordinates {(Top 6, 0.8) (Top 12, 0.73) (Top 18, 0.71) (Top 24, 0.74) (Top 30, 0.73)};
		
		\addplot +[
		black,
		fill=green!20,
		postaction={
			pattern=crosshatch
		}
		]
		coordinates {(Top 6, 0.93) (Top 12, 0.85) (Top 18, 0.84) (Top 24, 0.83) (Top 30, 0.83)};

		\legend{DNMF,GNMF,GSNMF}
		
		\end{groupplot}
		\end{tikzpicture}
	}
	\caption{The accuracy of community profiling methods in extracting relevant key expressions.}
	\label{fig:profiling_accuracy}
\end{figure*}

Table \ref{tab:UK-GNMF} shows the profiles of five communities detected by GNMF in the UK dataset as well as their corresponding ground truth political parties. Due to space limitation, we do not provide the community profiles detected by DNMF. As we observe from Table \ref{tab:UK-GNMF}, the same problem which exists in the profiles of communities detected by GNMF and DNMF in the US dataset still exists here. In other words, it is not clear which community represents which party. For instance, the profiles which corresponds to the Conservative Party and the Labour Party shared many key expressions such as Labour, Tories, David Cameron (@David\_Cameron), @Ed\_Miliband, UKLabour, and VoteLabour, but there is no other information to understand the positions of these two parties towards these key expressions in order to differentiate them and also associate the community profiles to the parties. However, as Table \ref{tab:UK-motives} suggests, the community profiles detected by our framework shows that the community associated to the Conservative Party have positive attitude towards David Cameron but negative attitude towards Labour and Miliband. On the other hand, the community associated to the Labour Party have the positive attitude towards Labour, UKLabour, and Miliband but negative attitude towards Tories and David Cameron. Since this corresponds to our ground truth, we can conclude that sentiment information can play an essential role in providing better community profiles.

\subsubsection{Quantitative results.}
\label{Quantitative_results}
In this section, we aim to compare GSNMF with GNMF and DNMF in terms of their effectiveness in extracting relevant key expressions for community profiles. Figure \ref{fig:profiling_accuracy} shows the accuracy of all methods in US and UK datasets by considering different number of top key expressions as community profiles. As we observe, GSNMF outweighs GNMF and DNMF in all experiments. For instance, by considering top 30 key expressions as community profiles, $93\%$ of key expressions extracted by GSNMF in the US dataset are relevant compared to $82\%$  in GNMF and $85\%$ in DNMF. Similarly, $83\%$ of key expressions extracted by GSNMF in the UK dataset are relevant compared to $65\%$ in DNMF and $73\%$ in GNMF. The experiments also suggest that GSNMF achieves better accuracy with lower number of top key expressions as community profiles. This implies that the higher a key expression is ranked by GSNMF, the more likely it is relevant. Following these observations, sentiment-driven community profiling produces key expressions which are more relevant than its sentiment insensitive counterparts.



\section{Conclusion and Future Work}
In this paper, we presented a sentiment-driven community profiling and detection framework. Our framework uncovers a low-dimensional latent space in order to cluster users according to their opinions and social interactions. It also provides community profiles reflecting positive/negative collective opinions of their members. Experimental results on real-world social media datasets demonstrated: (1) our framework obtains significant performance in detecting communities compared to several state-of-the-art community detection methods, and (2) our framework presents a sentiment-driven community profiling approach that provides better insights into the collective opinions of community members by dividing key expressions into positive and negative categories.


Our future work includes the following directions. First, the current sentiment analysis is not capable of differentiating between opposition and concern. There is a need to propose new methods to differentiate between opposition and concern. Second, identifying the dynamics of communities sheds light on their temporal behavior. Therefore, we will focus our efforts on detecting and profiling the dynamics of communities.

%
%
%
%
%
%

\appendix
\section{Appendix}
\subsection{Computation of U}
Optimizing the objective function $\mathcal{F}$ in Eq. \eqref{proposed_method} with respect to $\mathbf{U}$ is equivalent to solving

\begin{equation}
\begin{aligned}
\underset{\mathbf{U}}{\text{min}}\quad & \mathcal{F}_U = ||\mathbf{X} - \mathbf{V} \mathbf{U}^T ||_F^2 - \lambda Tr(\mathbf{U}^T \overset{\sim}{\mathbf{W}} \mathbf{U})\\
\text{s.t.}\quad & \mathbf{U} \geq 0, \mathbf{U}^T \mathbf{U} = \mathbf{I}.
\end{aligned}
\end{equation}

Let $\mathbf{\Gamma}$ and $\mathbf{\Lambda}$ be the Lagrange multiplier for constraints $\mathbf{U}^T \mathbf{U} = \mathbf{I}$ and $\mathbf{U} \geq 0$ respectively, and the Lagrange function is defined as follows:
\begin{equation}
\begin{aligned}
\underset{\mathbf{U}}{\text{min}} \quad & \mathcal{L}_\mathbf{U} = ||\mathbf{X} - \mathbf{V} \mathbf{U}^T ||_F^2 - \lambda Tr(\mathbf{U}^T \overset{\sim}{\mathbf{W}} \mathbf{U})\\
& \quad \quad - Tr(\mathbf{\Lambda} \mathbf{U}^T) + Tr(\mathbf{\Gamma} (\mathbf{U}^T \mathbf{U} - \mathbf{I}))
\end{aligned}
\end{equation}

The derivative of $\mathcal{L}_\mathbf{U}$ with respect to $\mathbf{U}$ is
\begin{equation}
\begin{aligned}
\frac{\partial \mathcal{L}_\mathbf{U}}{\partial \mathbf{U}} = -2 \mathbf{X}^T \mathbf{V} + 2 \mathbf{U} \mathbf{V}^T \mathbf{V} - 2 \lambda \overset{\sim}{\mathbf{W}} \mathbf{U} - \mathbf{\Lambda} + 2 \mathbf{U} \mathbf{\Gamma}
\end{aligned}
\end{equation}

By setting $\frac{\partial \mathcal{L}_\mathbf{U}}{\partial \mathbf{U}} = 0$, we get
\begin{equation}
\begin{aligned}
\mathbf{\Lambda} = -2 \mathbf{X}^T \mathbf{V} + 2 \mathbf{U} \mathbf{V}^T \mathbf{V} - 2 \lambda \overset{\sim}{\mathbf{W}} \mathbf{U} + 2 \mathbf{U} \mathbf{\Gamma}
\end{aligned}
\end{equation}

With the KKT complementary condition for the nonnegativity of $\mathbf{U}$, we have $\mathbf{\Lambda}_{ij} \mathbf{U}_{ij} = 0$. Therefore, we have
\begin{equation}
\begin{aligned}
(-\mathbf{X}^T \mathbf{V} + \mathbf{U} \mathbf{V}^T \mathbf{V} - \lambda \overset{\sim}{\mathbf{W}} \mathbf{U} + \mathbf{U} \mathbf{\Gamma})_{ij} \mathbf{U}_{ij} = 0
\end{aligned}
\end{equation}
where  $\mathbf{\Gamma} = \mathbf{U}^T \mathbf{X}^T \mathbf{V} - \mathbf{V}^T \mathbf{V} + \lambda \mathbf{U}^T \overset{\sim}{\mathbf{W}} \mathbf{U}$.

Matrices $\mathbf{\Gamma}$, $\mathbf{X}^T \mathbf{V} $, and $\mathbf{V}^T \mathbf{V}$  take mixed signs. Motivated by \cite{ding2010convex}, we separate positive and negative parts of any matrix $\mathbf{A}$ as $\mathbf{A}_{ij}^+ = (|\mathbf{A}_{ij}| + \mathbf{A}_{ij})/2, \quad  \mathbf{A}_{ij}^- = (|\mathbf{A}_{ij}| - \mathbf{A}_{ij})/2.$

Thus, we get
\begin{equation}
\begin{aligned}
[-((\mathbf{X}^T \mathbf{V})^+ + [\mathbf{U}(\mathbf{V}^T \mathbf{V})^-] + \lambda \overset{\sim}{\mathbf{W}} \mathbf{U} + \mathbf{U} \mathbf{\Gamma}^-)\\ + ((\mathbf{X}^T \mathbf{V})^- + [\mathbf{U}(\mathbf{V}^T \mathbf{V})^+] + \mathbf{U} \mathbf{\Gamma}^+)]_{ij} \mathbf{U}_{ij} = 0
\end{aligned}
\end{equation}

which leads to the updating rule of $\mathbf{U}$ in Eq \eqref{U_updating}. 

\subsection{Computation of V}

Optimizing the objective function $\mathcal{F}$ in Eq. \eqref{proposed_method} with respect to $\mathbf{V}$ is equivalent to solving

\begin{equation}
\begin{aligned}
\underset{\mathbf{V}}{\text{min}}\quad & \mathcal{F}_\mathbf{V} = ||\mathbf{X} - \mathbf{V} \mathbf{U}^T ||_F^2
\end{aligned}
\end{equation}

The derivative of $\mathcal{F}_\mathbf{V}$ with respect to $\mathbf{V}$ is
\begin{equation}
\label{V_deriv}
\begin{aligned}
\frac{\partial \mathcal{F}_\mathbf{V}}{\partial \mathbf{V}} = -2 \mathbf{X} \mathbf{U} + 2 \mathbf{V} \mathbf{U}^T \mathbf{U}
\end{aligned}
\end{equation}

Setting $\frac{\partial \mathcal{F}_\mathbf{V}}{\partial \mathbf{V}}=0$, we get the updating rule of $\mathbf{V}$ in Eq \eqref{V_updating}.

\bibliographystyle{ACM-Reference-Format}
\bibliography{references}

\end{document}